\documentclass[%
 aip,
 amsmath,amssymb,
 reprint,%
]{revtex4-1}

\usepackage{graphicx}
\usepackage{dcolumn}
\usepackage{bm}

\usepackage[utf8]{inputenc}
\usepackage[T1]{fontenc}
\usepackage{mathptmx}
\usepackage{etoolbox}
\usepackage{xcolor}
\usepackage{braket}

\makeatletter
\def\@email#1#2{%
 \endgroup
 \patchcmd{\titleblock@produce}
  {\frontmatter@RRAPformat}
  {\frontmatter@RRAPformat{\produce@RRAP{*#1\href{mailto:#2}{#2}}}\frontmatter@RRAPformat}
  {}{}
}%
\makeatother

\begin{document}

\preprint{AIP/123-QED}

\title[The role of interaction in matter wave optics]{The role of interaction in matter wave optics with motional states}
\author{RuGway Wu}
\affiliation{Vienna Center for Quantum Science and Technology, Atominstitut, TU Wien, Stadionallee 2, 1020 Vienna, Austria}

\author{Maximilian Pr\"{u}fer}
\affiliation{Vienna Center for Quantum Science and Technology, Atominstitut, TU Wien, Stadionallee 2, 1020 Vienna, Austria}

\author{J\"{o}rg Schmiedmayer}
\email{schmiedmayer@atomchip.org}
\homepage{http://www.atomchip.org}
\affiliation{Vienna Center for Quantum Science and Technology, Atominstitut, TU Wien, Stadionallee 2, 1020 Vienna, Austria}

\date{\today}

\begin{abstract}

Matter-wave optics is often viewed as a linear analogue of photonics, where noninteracting particles are coherently split, diffracted, and recombined, and interference arises from single-particle coherence. In ultracold quantum gases, however, interactions are intrinsic and can rival or exceed kinetic and optical energy scales. This drives matter-wave optics into a nonlinear regime: diffraction and momentum distributions become interaction-dependent, interference contrast degrades or collapses, and revival dynamics appear. In the mean time, interactions can generate squeezing and entanglement, enabling sensitivities beyond the standard quantum limit. We showcase representative examples—covering diffraction, splitting, and interferometry—that illustrate how interactions reshape the basic elements of matter-wave optics and open new opportunities for nonlinear quantum technologies.

\end{abstract}

\maketitle

\tableofcontents

\section{Introduction}

The wave nature of matter, first proposed by Louis de Broglie a century ago and spectacularly confirmed by electron diffraction \cite{ Davisson_1927_electron_diffraction} and later by neutron interferometry \cite{ Rauch_neutron_interferometry_book}, is one of the foundation of modern quantum physics. The realization that massive particles can exhibit interference phenomena established the discipline of matter-wave optics. Early demonstrations with neutrons showed that even composite massive particles behave coherently under well-controlled conditions, inspiring analogous studies with atomic and molecular beams where both internal and external degrees of freedom can be precisely controlled \cite{Ramsey_1996_beam}.

The advent of laser cooling and trapping \cite{Phillips_1985_cooling_trapping} revolutionized atomic physics, transforming individual atoms from high-velocity projectiles into controllable quantum objects and paving the way for a wide range of studies in coherent atom optics. The subsequent realization of Bose–Einstein condensation (BEC) in dilute gases \cite{Anderson1995,Davis1995,Bradley1995} provided an unprecedented source of coherent matter waves. The observation of interference between independently prepared condensates confirmed that matter waves can display macroscopic coherence comparable to optical laser fields \cite{Andrews1997}. A comprehensive overview of optics and interfereometry with atoms and molecules is provided in \cite{CroninRMP2009}.

Yet, unlike photons, atoms inevitably interact. At ultralow temperatures, binary s-wave collisions result in mean-field energy shifts that can be comparable to (or even larger then) the kinetic and optical energy scales that govern diffraction and interferometry. As a result, atom optics cannot remain purely linear: diffraction patterns are reshaped, interference phases become density-dependent, and coherence can be lost through interaction-induced dephasing \cite{Lewenstein1996,Castin1997,Sinatra2001}. Conversely—and in direct analogy to nonlinear optics—the same interactions also enable new phenomena such as squeezing and the suppression of quantum noise \cite{Kitagawa_1993_squeezed_spin_states,Riedel2010,Gross_2010_nonlinear_atom_interferometer,PezzeSmerziRMP2018}. 

The strength and even the sign of atomic interactions can be tuned using magnetic or optical Feshbach resonances \cite{ChinRMP2010}, granting experimental access from weakly to strongly interacting regimes. Together, these features transform matter-wave optics into a nonlinear discipline in which interactions are not merely a limitation, but a resource for engineering squeezing, entanglement, and nonlinear interferometric response.

In this perspective, we do not attempt to provide a comprehensive review of nonlinear matter-wave optics; rather, we highlight several representative examples that illustrate how interparticle interactions modify the fundamental processes of diffraction, splitting, and interference. These examples are mostly drawn from our own experiments and are meant to place recent progress in a broader context.

We first examine the diffraction of strongly interacting Feshbach molecules, where the interplay between the optical lattice potential and strong molecular interactions leads to coherent yet slowed diffraction dynamics, revealing signatures of beyond-mean-field behavior.

We then discuss the coherent splitting of a Bose–Einstein condensate in a double-well potential, which serves as a matter-wave beam splitter and further enables matterwave interferometry. Here, interactions simultaneously enable the creation of squeezed and entangled states—resources for quantum-enhanced metrology—while also introducing dephasing and decoherence that limit coherence times. The splitting process inherently drives the system out of equilibrium, and matter-wave interference provides a powerful probe of interaction-driven non-equilibrium quantum evolution.

Finally, we turn to interferometry with trapped Feshbach molecules, where both internal and motional degrees of freedom participate in interference. In this strongly interacting regime, phase shifts, contrast, and coherence are all governed by nonlinear dynamics, demonstrating how even the simplest optical elements become intrinsically nonlinear when applied to interacting matter waves.

Together, these examples trace a coherent theme: Interaction-driven nonlinear effects in matter-wave optics have evolved from an unwanted complication into a versatile tool—one that allows us to engineer, probe, and ultimately exploit a new degree of control in quantum science.

\paragraph*{Dedication.} 
This article is dedicated to Ernst M.~Rasel on the occasion of his 60th birthday. 
His pioneering contributions to matter-wave interferometry, atom optics, and precision measurements have shaped the field for more than three decades and continue to inspire its ongoing developments.

\section{Diffraction of strongly interacting molecules} \label{section:Li_molecule_diffraction}

Having established the broader context of interactions marking the transition from linear to nonlinear matter-wave optics, we begin with its most elementary building block: diffraction. Just as diffraction from gratings defines the foundations of classical and optical interferometry, it also provides the simplest setting to explore how interactions modify coherent evolution in matter waves.

As light can be split by diffraction from a material grating, matter waves can be split using gratings of light. Diffraction occurs when the phase of a wave gets modulated locally. When a wave encounters a periodic structure, such as when atoms encounter a standing wave potential $V(x)=V_0 \cos^2(k_L x)$ formed by the interference of coherent light \cite{kapitza_dirac_1933}, this produces discrete output modes of distinct momenta. The phenomenon is generally treated as resulting from the coherent superposition and interference of waves propagating in different paths through the diffracting region that have the same starting and ending points. It can also be understood as two-photon scattering processes transferring momenta in units of the vector sum of the two photons recoils to a particle. As a result, diffraction (or the closely related Raman transitions) split the input state into a coherent superposition of momentum states. Since the first demonstrations of the diffraction of atoms from an optical standing wave \cite{Moskowitz_1983_atom_diffraction_standing_wave} and from a material grating \cite{Keith_1988_atom_diffraction_grating}, its application has led to many interference experiments with atoms and molecules \cite{CroninRMP2009}. Early experiments with weakly interacting atomic BECs studied both the short pulse Ramn--Nath and the Bragg regimes, and their crossovers~\cite{Kozuma1999,Ovchinnikov1999,Deng1999,Campbell_2005_photon_recoil,Gadway_2009_KD_beyond_RamanNath}.

\begin{figure}[t]
\includegraphics[width=1\columnwidth]{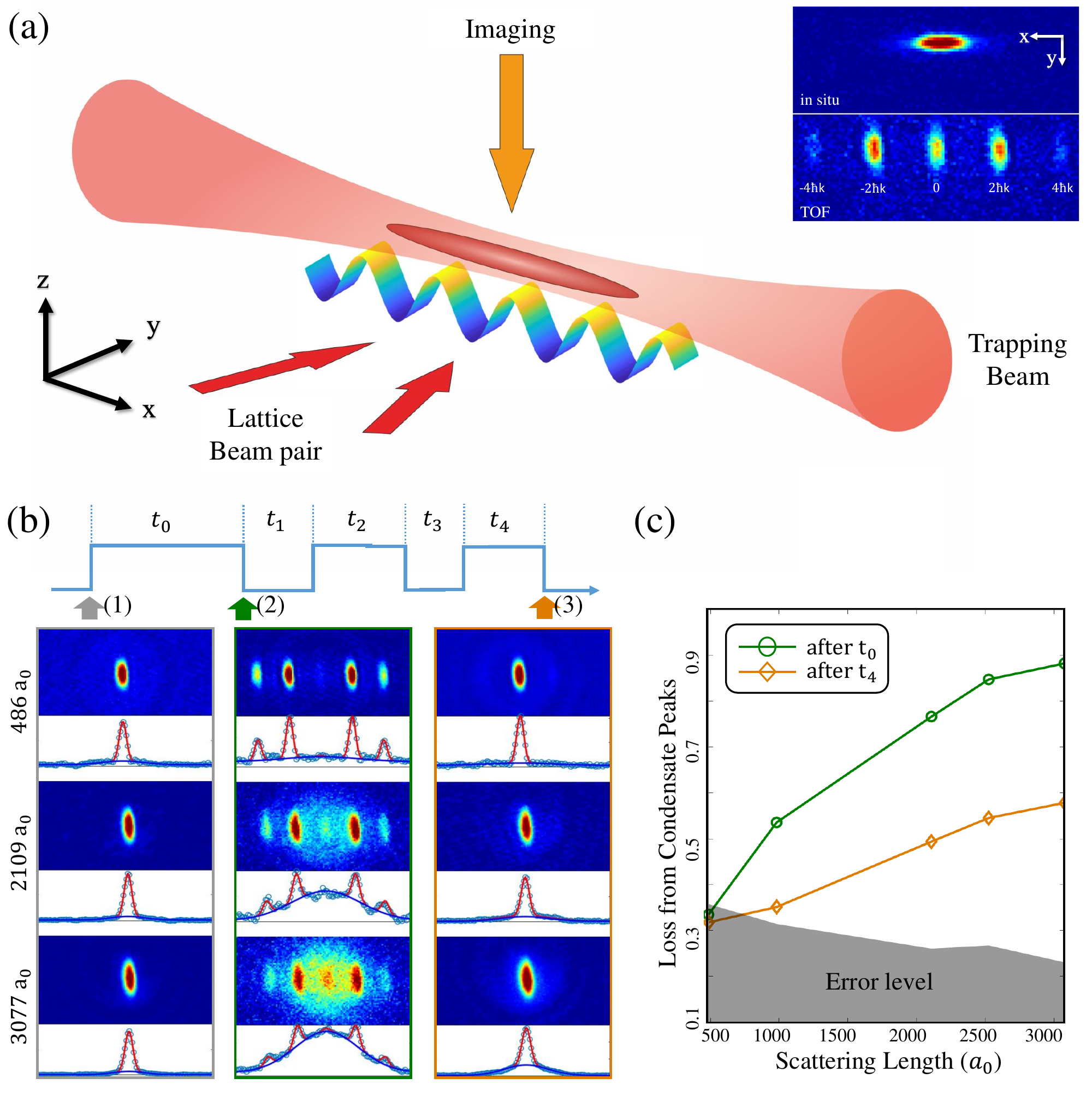}
\caption{\label{fig:diffraction_setup} \textit{(a)} Schematic of the diffraction of strongly interacting mBECs from an optical lattice, with momentum space images taken before and after matter-wave focusing. Using a designed pulse sequence to return the diffracted molecules to the condensate \textit{(b)} reveals the sources of scattering loss \textit{(c)}. Figure adapted from \cite{Liang2022}.}
\end{figure}

\begin{figure*}
\includegraphics[width=2\columnwidth]{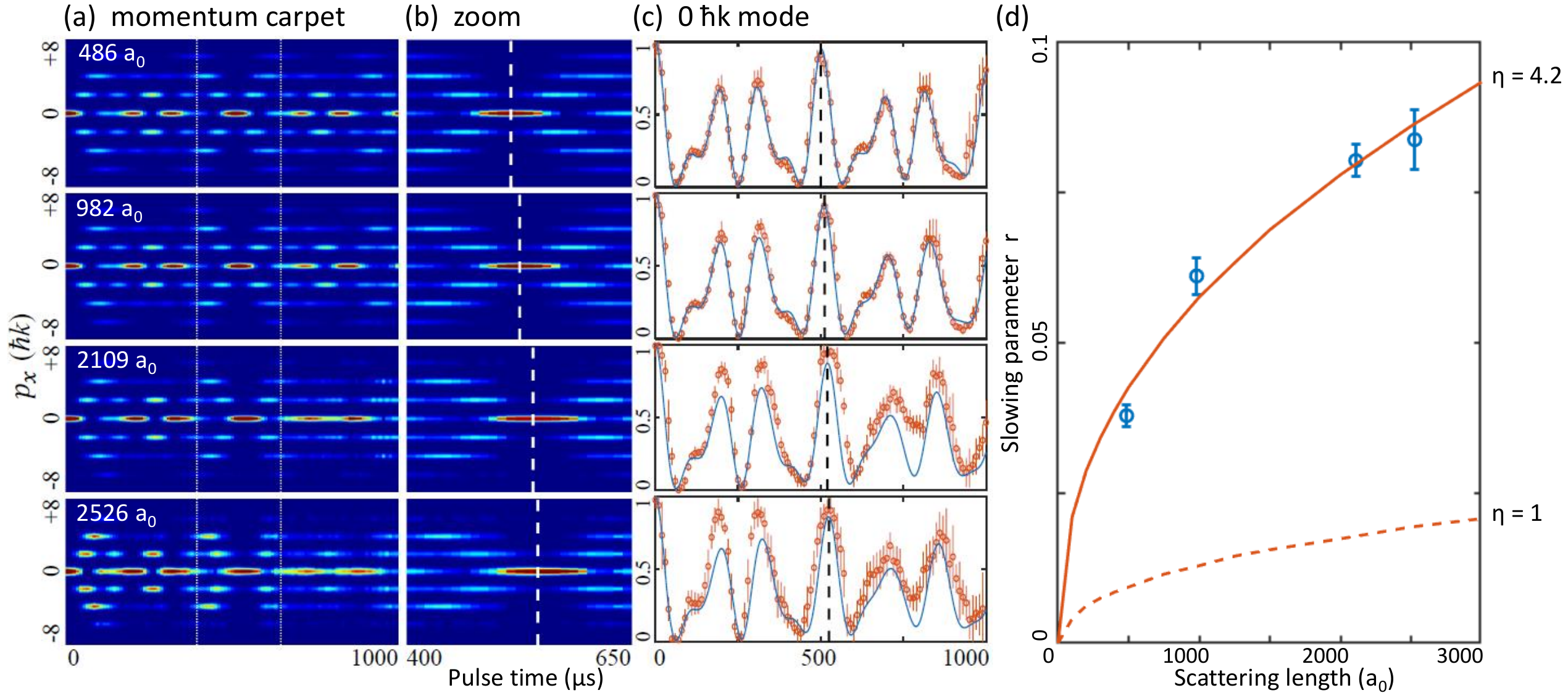}
\caption{\textit{(a)} Time carpets of momentum space distribution during the diffraction process for increasing interaction strengths (with scattering background removed for better visualization), showing a slowing-down of the evolution. \textit{(b)} Zoomed view of the condensate recurrence peak. \textit{(c)} The corresponding normalized populations of the 0 $\hbar k$ momentum mode (circle), and 1D GPE simulation results including a phenomenological scaling factor $\eta = 4.2$ (solid curve). \textit{(c)} Slowing parameter $r$ for quantifying the slowing down from experimental data, and simulation results with and without scaling the interaction strength by the fudge factor $\eta$. Figure adapted from \cite{Liang2022}.}
\label{fig:diffraction_carpet}
\end{figure*}

The extension of atom-optical techniques into regimes of strong interparticle interaction requires a detailed understanding of how interactions modify both the intrinsic many-body dynamics and the coupling between matter waves and light fields. 
To explore these effects, diffraction of a strongly interacting molecular Bose--Einstein condensate (mBEC) from optical standing-wave pulses was investigated using lithium Feshbach molecules~\cite{Liang2022}. 

The fermionic isotope of lithium, $^{6}$Li, provides an ideal platform for studying strongly interacting quantum gases. 
Near the broad Feshbach resonance between two spin states, weakly bound Li$_2$ molecules can be created with a tunable $s$-wave scattering length $a_s$ spanning thousands of Bohr radii~\cite{ChinRMP2010}. 
Because of the fermionic nature of the constituent atoms, inelastic losses are strongly suppressed at large positive scattering length~\cite{Petrov_2004_dimer,Petrov_2005_dimer_scattering}. 
This makes $^{6}$Li$_2$ an exceptionally stable system of strongly interacting bosons, ideally suited for investigating coherent matter-wave dynamics in the presence of strong interactions.

As qualitatively depicted in Figure~\ref{fig:diffraction_setup} \textit{(a)}, the mBEC confined in a trapping potential is subject to pulses of a lattice potential for variable times. The interaction energy of the condensate, characterized by the chemical potential, is comparable to the single photon recoil energy of the lattice, $E_r=\hbar^2 k^2/2m \approx 250\,\mathrm{Hz}$, where $k = \pi/D$ and $m$ is the mass of a lithium molecule ($^6Li_2$). Diffraction of the strongly interacting mBEC from the optical grating leads to a number of key observations~\cite{Liang2022}: 
Key observations include (i) \emph{coherent diffraction processes} from the optical grating in the presence of strong microscopic molecule-molecule interaction, but with
(ii) \emph{reduced visibility} due to incoherent collisions within the sample during ther separation of the diffraction orders; and 
(iii) \emph{slowing down of the momentum mode evolution} which depends on the strength of interaction.

\subsection{Molecule-molecule scattering}

In the presence of strong interaction, diffraction nonetheless gives rise to coherent evolution of the momentum modes. While distinct momentum modes can be recognized from the diffraction patterns (Figure \ref{fig:diffraction_setup} \textit{(b)}), an enhanced presence of the broad background with increased interaction indicates an interaction-dependent loss of molecules from the condensate. 
Following an initial diffraction pulse, the populations in the diffracted non-zero momentum modes can be predominantly returned to the 0 $\hbar k$ condensate by applying a subsequent designed pulse sequence \cite{Zhou_2018_shortcut_loading} before the momentum modes spatially separate, consequently diminishing the scattering background. This clearly demonstrates the coherence of the diffraction processes, and reveals two principal contributions to the interaction-dependent degradation of the coherent diffraction patterns (Figure \ref{fig:diffraction_setup} \textit{(c)}): (i) collisions between different momentum modes during separation, and (ii) in-trap loss of molecules during the lattice pulse, which includes dissociation of Feshbach molecules into unbound atoms, as confirmed by radio-frequency spectroscopy.

\subsection{Interaction-induced slowing down}

Fitting and then removing the incoherent background produced by incoherent molecule-molecule scatttering and renormalizing the momentum mode populations reveals clearly the evolution of the coherent modes. For short strong pulses, the standard Kapitza-Dirac diffraction in the Raman-Nath regime is obtained, which matches very closely with theoretical predictions. 

For longer pulses with $U_0\gg E_r$, known as the channeling regime\cite{keller_adiabatic_1999}, particles oscillate within each lattice site, creating a periodic pattern in momentum space, such as shown in Figure \ref{fig:diffraction_carpet}. The effect of interaction is clearly revealed at longer times: the time point where the coherent population is restored to the $0\hbar k$ mode (recurrence of the mBEC) occurs at increasingly later time points in the presence of stronger interaction (Figure \ref{fig:diffraction_carpet}) \textit{(c)}), showing that the evolution becomes slower.

A simple physical picture is proposed for the slowing down of the scattering process: A Density grating forms across the condensate in the presence of the lattice potential, as the higher momentum mode populations grow. The effect of repulsive interaction counters the effect of the lattice, slowing down the evolution process. The interaction-induced slowing effect itself is reproduced by full 3D as well as 1D mean-field GPE simulations, confirming that this phenomenon occurs along the lattice direction and is well captured by a 1D model. However, simulations performed with experimental parameters and the scattering lengths at respective magnetic fields generate evolutions with much weaker slowing down than observed experimentally. Quantitatively matching the observed slowing requires an enhancement of $a_s$ by including a phenomenological fudge factor $\sim\!4$ in numerical simulations. This indicates the presence of additional interaction energy contributing to the effect, suggesting beyond-mean-field physics and/or correction to theory at very strong interactions.

\subsection{Open questions}

The observed diffraction processes in the Raman--Nath and channeling regimes allow for a direct comparison of interaction effects. When the action of an ``optical element’’ is fast, as in the Raman--Nath regime, interaction effects can often be neglected. Despite the significant collisional loss that occurs once the matter-wave packets begin to separate, diffraction from a strong optical standing wave remains largely coherent, and the population dynamics agree quantitatively with theoretical predictions.

In contrast, if the optical standing wave remains on for times long compared to the collision time, the evolution clearly reveals the nonlinearity arising from interparticle interactions.  The experiments in a regime where the strongly interacting Li\(_2\) molecules channel in an optical grating (lattice)  therefore provide an example where interactions are non-negligible and substantially modify the observed diffraction dynamics. In this sense, the optical grating acts as a \emph{nonlinear optical element} for interacting matter waves. Here, the mean-field interaction plays a role analogous to Kerr nonlinearities and self-phase modulation in photonics: the phase evolution acquires a density-dependent contribution that directly alters the diffraction pattern.

As a related development, it has been demonstrated recently how light scattering can be modified by interatomic interaction \cite{Konstantinou_2025_boson_light_scattering} and spatial pattern of the atoms \cite{Lu_2025_suppression_light_scattering}, offering some further insight for the collective response of atoms to optical fields. Bosonic stimulation of light scattering rate is given by the structure factor \cite{Lu_2022_bosonic_stimulation}, while repulsive interaction results in a change of short-distance atom-atom correlation, giving rise to beyond mean-field effect suppressing wave function overlap and the subsequent enhancement \cite{Konstantinou_2025_boson_light_scattering}. Coherent light scattering from highly ordered atoms in periodic arrays exhibit constructive and destructive interference depending on the angles of the incident and diffracted light, with defects or disorder reducing the degree of light scattering modification \cite{Lu_2025_suppression_light_scattering}. If such insights can be generalized from photon scattering to matter-wave processes, they may offer a promising route toward resolving the observed “fudge-factor puzzle’’ in nonlinear matter-wave optics, where the experiments point to a modified effective interaction strengths.

Incoherent collision loss plays a significant role in matter-wave optics under strong interaction, leading to reduction of signal. Two-body collisions are also the root of interparticle interaction. To date, there has been very few experimental studies of collisional scattering processes in matter wave optical elements. In the recent work by Chen et al.\cite{Chen_2025_scattering_halo}, the scattering between Li$_2$ molecules as two overlapping wave packets separate, the same scenario as encountered as in the diffraction experiment \cite{Liang2022}, is examined more closely. This work also demonstrates deviation from perturbation theory, which necessitates theoretical refinements to account for nonlinear effects and many-body correlations.

A unified description that captures both coherent nonlinear evolution and interaction-induced decoherence will therefore be essential for advancing nonlinear matter-wave devices and for achieving quantitative control in strongly interacting quantum gases.

\section{Coherently splitting a BEC in a double well} \label{section:coherent_splitting}

So far, we considered momentum modes as the output ports of our atomic beam splitter. Another possibility is to use two spatial modes, such as in a double-well potential. The idea is to split a single BEC into two BECs residing on both sides of a double well, which are the output ports of this atomic beamsplitter. Two immediate questions arise: 1) What is the second, potentially empty input port? And 2) What is the role of interactions during the splitting process? In the following, we will discuss this type of atomic beam splitter and give examples of experimental implementations and applications of double well beam splitters.

\subsection{Beamsplitter by spatial splitting}

\begin{figure}
\includegraphics[width=\columnwidth]{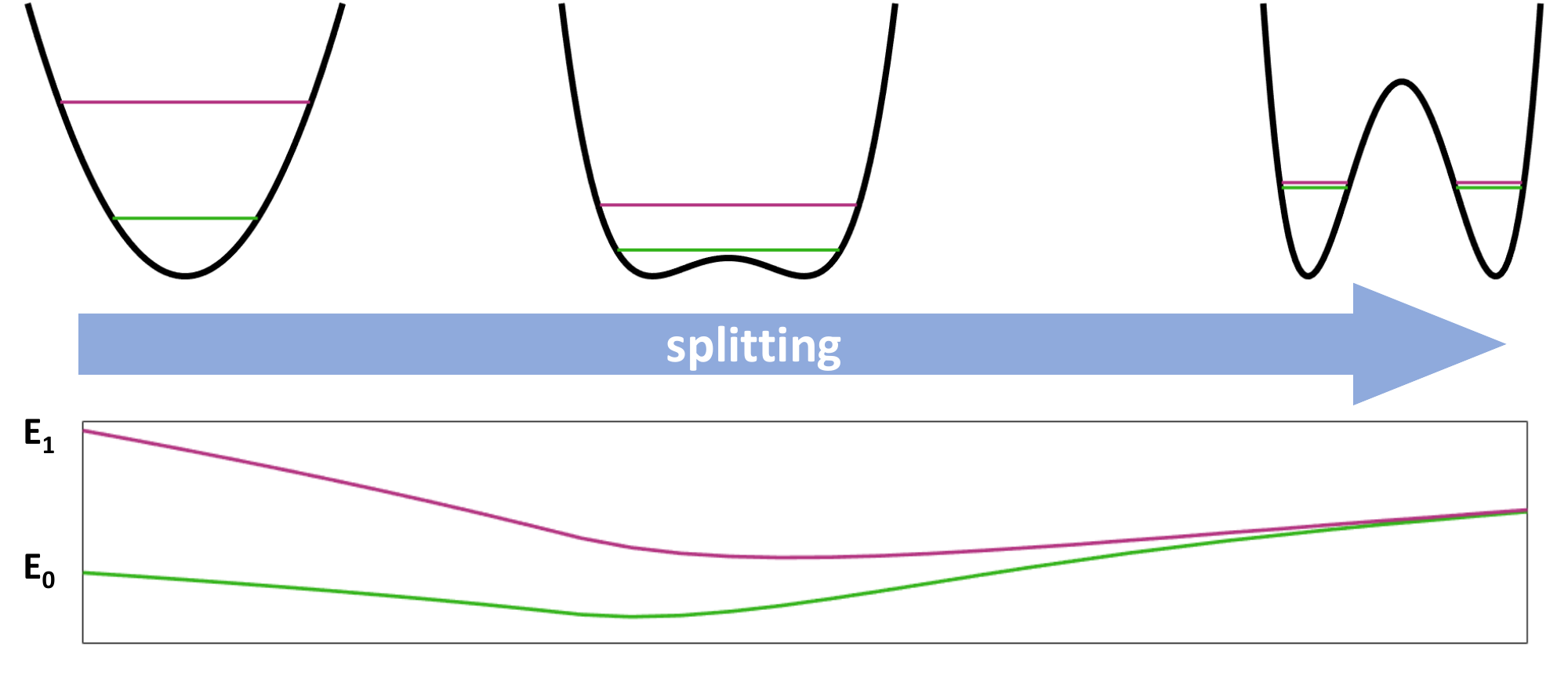}
\caption{Splitting process, showing the ground state and first excited state energy levels. A BEC is prepared in a single well. The chemical potential energy and the temperature are lower than the harmonic oscillator energy such that the first exited state is empty and only features quantum fluctuations; it is the empty port of the beam splitter. For a fully split double well the first two eigenstates become degenerate and one finds a good description in a left/right basis (the two output ports of the beamsplitter).}
\label{fig:schematicsplitting}
\end{figure}

To introduce the basic operation of a double-well beam splitter, we consider the process in which a single-well trap is smoothly deformed into a double-well potential (see Fig.\,\ref{fig:schematicsplitting}).  In the absence of interactions, the initial ground state evolves adiabatically into the symmetric superposition of the two wells, while the first excited state of the initial trap plays the role of the empty input port of a conventional optical beam splitter. This operation therefore realizes a coherent 50/50 splitting of the matter wave: the atoms populate the two wells on average with equal occupation, with zero mean number imbalance, and the corresponding fluctuations are at the shot-noise level. 

To investigate the effects of interactions, we consider a Bose–Einstein condensate prepared in a single-well potential characterized by a chemical potential $\mu$ and temperature $T$. If both quantities are much smaller than the harmonic confinement along the splitting direction, $T,\mu \ll \hbar\omega_x$, the condensate remains in the lowest trapped state. Adiabatically transforming this single-well trap into a double-well (Fig.\,\ref{fig:schematicsplitting}) dynamically splits the condensate into two spatially separated parts. For a fully separated double well, the two lowest eigenstates become nearly degenerate, justifying a description in terms of left- and right-localized modes. 

In the presence of interparticle interactions, the splitting process itself can generate quantum correlations between the two emerging wells. As the potential barrier is raised, repulsive interactions penalize population imbalance and thereby produce relative-number squeezing between the two modes. The degree of squeezing depends sensitively on the speed of the splitting: slow, near-adiabatic ramps allow interactions to suppress number fluctuations below the shot-noise level, whereas fast splitting yields fluctuations close to those of a coherent state. The resulting quantum correlations provide a resource for quantum-enhanced interferometry and, in principle, enable sensitivities beyond the standard quantum limit~\cite{PezzeSmerziRMP2018}.

A tunnel-coupled Bose–Einstein condensate (BEC) provides a versatile platform for exploring macroscopic quantum dynamics in a controlled setting. 
When a dilute gas of bosons is confined in a double-well potential at ultracold temperatures, the system can be effectively described by two weakly linked condensates, each localized in one of the wells. 
Coherent tunneling between the two spatial modes gives rise to a \emph{bosonic Josephson junction} (BJJ)~\cite{Smerzi1997_Josephson}, in which the population imbalance and the relative phase form a pair of canonically conjugate variables, analogous to the supercurrent and phase difference in a superconducting Josephson junction~\cite{Golubov2004}. 
This makes the BJJ an ideal testbed for studying nonlinear quantum dynamics, interaction-driven phase evolution,  and the emergence of macroscopic coherence.

\subsection{Influence of dimensionality}
\label{section:dimensionality}

The double-well potential introduces an additional two-level degree of freedom along the transverse direction, effectively allowing particles to occupy either of the two wells. 
Depending on the confinement along the remaining spatial directions, the system can realize different effective dimensionalities, each giving rise to distinct coherence and dynamical properties.

\paragraph*{Three-dimensional regime.}

In a fully three-dimensional Bose-Einstein condensate the healing length is typically small compared to the ground state size of the trapping potential, and the chemical potential large compared to the confinement energy.
The condensate can be described within the mean-field Gross–Pitaevskii picture and the condensate is well described by a macroscopic wavefunction with weak phase fluctuations. In fact the first interference experiments with BEC were performed in a 3D double well \cite{Andrews1997}. The two-mode approximation is sufficient, splitting is robust as long the barrier is raised sufficiently slow, coherence is generally robust, and phase diffusion occurs only over long timescales. 
Many early demonstrations of matter-wave beam splitters and Josephson dynamics were performed in this regime \cite{Andrews1997,Shin2004_DoubleWell,Shin2005_SplitBEC,Jo2007_Recombination}.

\paragraph*{Zero-dimensional regime.}
The conceptually cleanest and experimentally most controlled realization of a double-well beam splitter is achieved under strong confinement, such that both the chemical potential and the temperature satisfy $\mu, k_B T \ll \hbar\omega_\perp$.  
In this zero-dimensional (0D) regime, the system is frozen into the transverse ground state, leaving only the tunneling axis as an active degree of freedom. 
The condensate behaves as an effective two-mode system, and the dynamics are governed by the bosonic Josephson-junction Hamiltonian~\cite{Smerzi1997_Josephson}. 
This regime enabled the first experiments demonstrating Josephson oscillations~\cite{Albiez2005}, and interaction-induced number squeezing and entanglement between the resultant BECs in the separate wells~\cite{Esteve2008,Gross_2010_nonlinear_atom_interferometer}.

\paragraph*{One-dimensional regime.}
Extending the 0D scenario by relaxing the confinement along one longitudinal direction leads to one-dimensional (1D) Bose gases~\cite{Krueger2010_1Dlimit} which can be split in the orthogonal direction into two parallel elongated condensates. 
This geometry, for example implemented on an atom chip~\cite{Schumm2005}, opened access to the rich field of relaxation and coherence dynamics beyond two-modes. Moreover, when the tunnel coupling is finite, the relative phase field of the two split 1D gases realizes the quantum sine–Gordon model~\cite{Gritsev2007,Schweigler_2017_correlations}, enabling quantitative studies of quantum field dynamics and tunneling phenomena~\cite{Zhang2024}. 
Thus, the 1D double-well system constitutes a powerful quantum simulator in addition to being an interferometric tool.

\paragraph*{Two-dimensional regime.}
In two dimensions, coupled condensate layers exhibit additional collective phenomena, allowing for example the exploration of Josephson dynamics in planar systems, including through the BEC–BCS crossover in ultracold Fermi gases~\cite{Valtolina_2015_Josephson_BEC_BCS,Luick_2020_2D_Fermi_Josephson}. 
For bosons, the interplay between tunneling and Berezinskii–Kosterlitz–Thouless (BKT) physics becomes relevant, with experiments revealing the role of vortex–antivortex excitations and algebraic order~\cite{Sunami_2022_BKT}. 

These studies highlight how dimensionality strongly influences coherence, correlations, and the emergent nonlinear dynamics of tunnel-coupled quantum gases.

\subsection{Experimental implementations.}

Experiments on matter-wave splitting can be carried out in various trapping configurations, such as optical or magnetic traps. Here, we focus on one particularly powerful experimental platform: the atom chip \cite{Folman_2002_atomchip}. Atom chips provide stable and tightly confining magnetic potentials that enable precise control over ultracold atomic ensembles. Moreover, a versatile and precisely controllable double-well potential can be realized through radio-frequency (RF) dressing of the magnetic trap~\cite{hofferberth2006radiofrequency,Lesanovsky2006,Lesanovsky2006a}. The double well can be well described by a very simple effective potential: \( \; \; V(x, \mathcal{A}) = a_2(\mathcal{A}) x^2 + a_4(\mathcal{A}) x^4 \; \; \) 
where the coefficients $a_2$ and $a_4$ are functions of the control parameter $\mathcal{A}$, proportional to the RF-current and reflect the curvature and separation of the wells \cite{wuerkner2025identificationoptimalcontrolstrategies}. $a_2$ changes sign during the splitting and $a_4$ stays nearly constant. Strikingly, this simple model captures all relevant physical features of the system with sufficient accuracy even for for optimal control purposes \cite{Kuriatnikov_2025_fast_splitting}.

Splitting of Bose–Einstein condensates (BECs) has been demonstrated in effectively one-dimensional systems, where it was shown that squeezing can be generated under appropriate experimental conditions \cite{Berrada2013}. Traditionally, the splitting speed has been constrained by experimental limitations: if performed too quickly, motional excitations are induced; if too slowly, decoherence from dephasing becomes significant. Recent studies have explored optimal control techniques (OCT) to achieve fast coherent and stable splitting \cite{Kuriatnikov_2025_fast_splitting} and two-step splitting protocols highlight that careful control of the dynamical evolution during the splitting process can be used to engineer highly entangled states \cite{Zhang2024}.

To emulate the behavior of a linear beam splitter, interactions during the splitting must be minimized. This can, in principle, be achieved by tuning the atomic interactions to zero via a Feshbach resonance—a technique feasible primarily in optical dipole traps \cite{Petrucciani_2025_mach_zehnder_noninteracting}. On atom chips, however, new approaches have emerged: fast, coherent splitting using optimal control techniques, suppressing motional excitations, has recently been demonstrated \cite{Kuriatnikov_2025_fast_splitting} (Fig. 4). It was shown that one has full control over the splitting time until close to the ultimate speed limit imposed by the finite trapping frequencies. Moreover calculations show that the capability to control the splitting dynamics will enable the generation of a highly squeezed and entangled states of external degrees of freedom \cite{Grond2009_OptimalControl}.

\begin{figure}
    \centering
    \includegraphics[width=\linewidth]{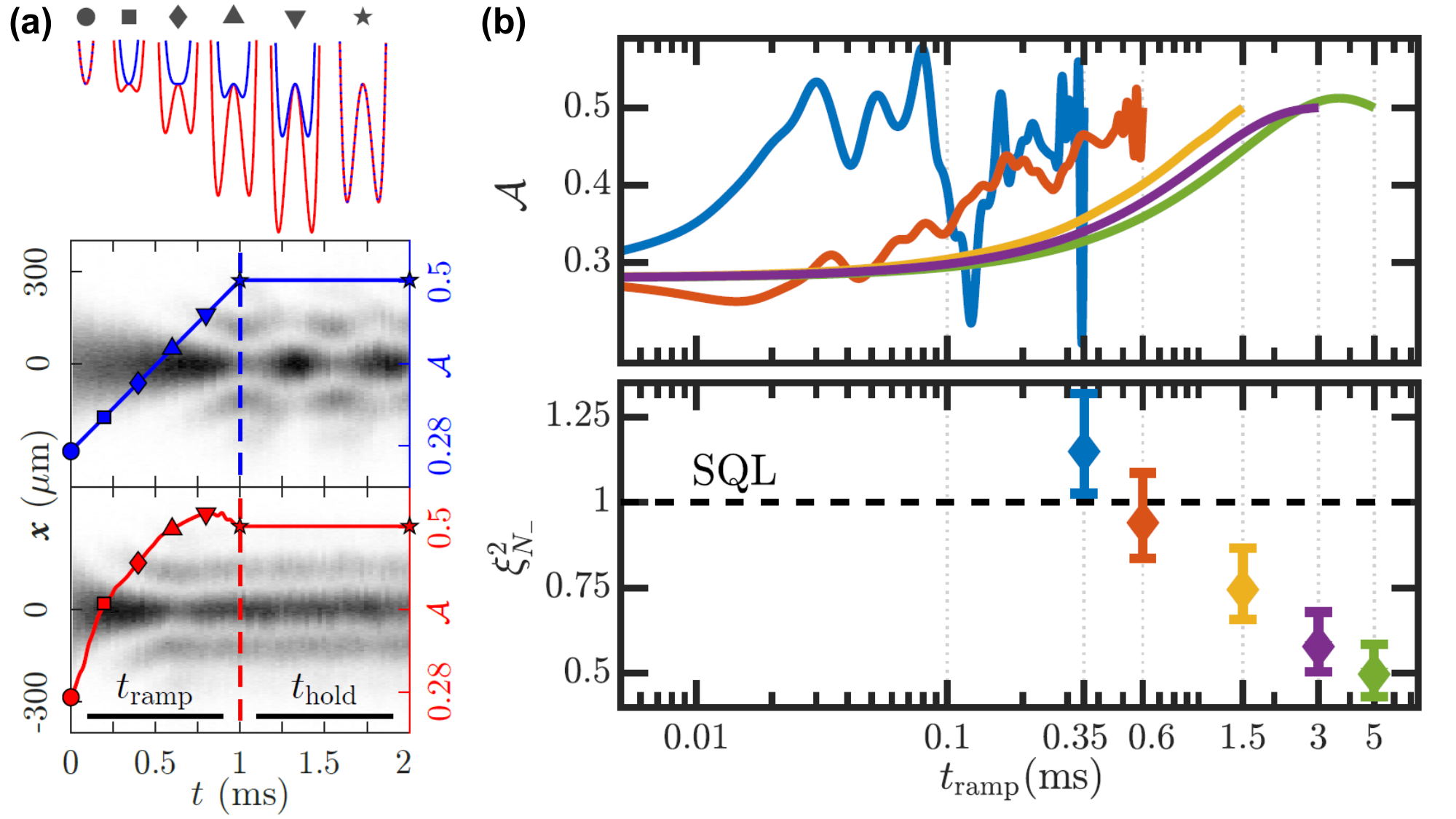}
    \caption{(a) Evolution of the interference pattern during ramp and hold times. Linear ramps (blue) induce strong excitations, whereas OC ramps (red) achieve splitting with negligible excitation. (b) Fast OC ramps to A = 0.5 trap with different ramp times (solid lines, upper panel), and the inferred number-squeezing factor (lower panel). Decreasing the ramp duration reduces the squeezing toward the standard quantum limit (SQL), as interaction does not result in significant effects in the short evolution time. Even for the fastest ramps, quantum properties are preserved, i.e., the relative phase is well-defined. Taken and adapted from \cite{Kuriatnikov_2025_fast_splitting}.}
    \label{fig:fast_splitting_OCT}
\end{figure}

\subsection{Implementing an interferometer after splitting}

There are two principal approaches to implementing an interferometer following the coherent splitting of a Bose--Einstein condensate (BEC) on an atom chip: the \emph{double-well interference scheme} and the \emph{Mach--Zehnder--type configuration}. Both rely on the coherent division of a condensate into two spatially separated parts and the subsequent interrogation of their relative phase evolution. However, they operate in different regimes and offer complementary insights into phase coherence, many-body dynamics, and interaction effects.
\subsubsection{Double-well interference}
In the {double-well interference scheme}, the two condensates are released and allowed to overlap during time-of-flight, producing an interference pattern that directly reflects the relative phase accumulated during the trapped evolution~\cite{Schumm2005}. The phase evolution is influenced by atom--atom interactions, which induce dephasing through the shot noise of the beam-splitting process. Although repulsive interactions can be employed to generate number-squeezed states and thereby reduce this dephasing, a residual phase diffusion remains, ultimately limiting the sensitivity of interaction-based interferometers~\cite{Grond2010_AtomInterferometry}.

Beyond their use for precision sensing, double-well systems constitute a powerful platform for exploring \emph{non-equilibrium quantum dynamics}. The act of splitting a coherent condensate represents a well-controlled quantum quench, launching the evolution of the relative phase and population imbalance between the two wells. In one-dimensional geometries, local interference fringes reveal the spatial dependence of this phase evolution, granting access to position-resolved coherence and correlation dynamics. Time-resolved measurements of these interference patterns have enabled the observation of interaction-driven dephasing~\cite{Hofferberth2007}, light-cone-like spreading of correlations~\cite{Langen2013}, pre-thermalization~\cite{Gring2012}, the demonstration of a Generalized Gibbs Ensemble~\cite{Langen2015}, many-body revivals~\cite{Rauer2018}, and many more.  Thus, the double-well configuration evolves from a static beam splitter into a tunable interferometric platform for probing interaction-driven relaxation and quantum dynamics in isolated many body systems.

\subsubsection{Mach-Zehnder interferometer}

\begin{figure}
    \centering
    \includegraphics[width=\linewidth]{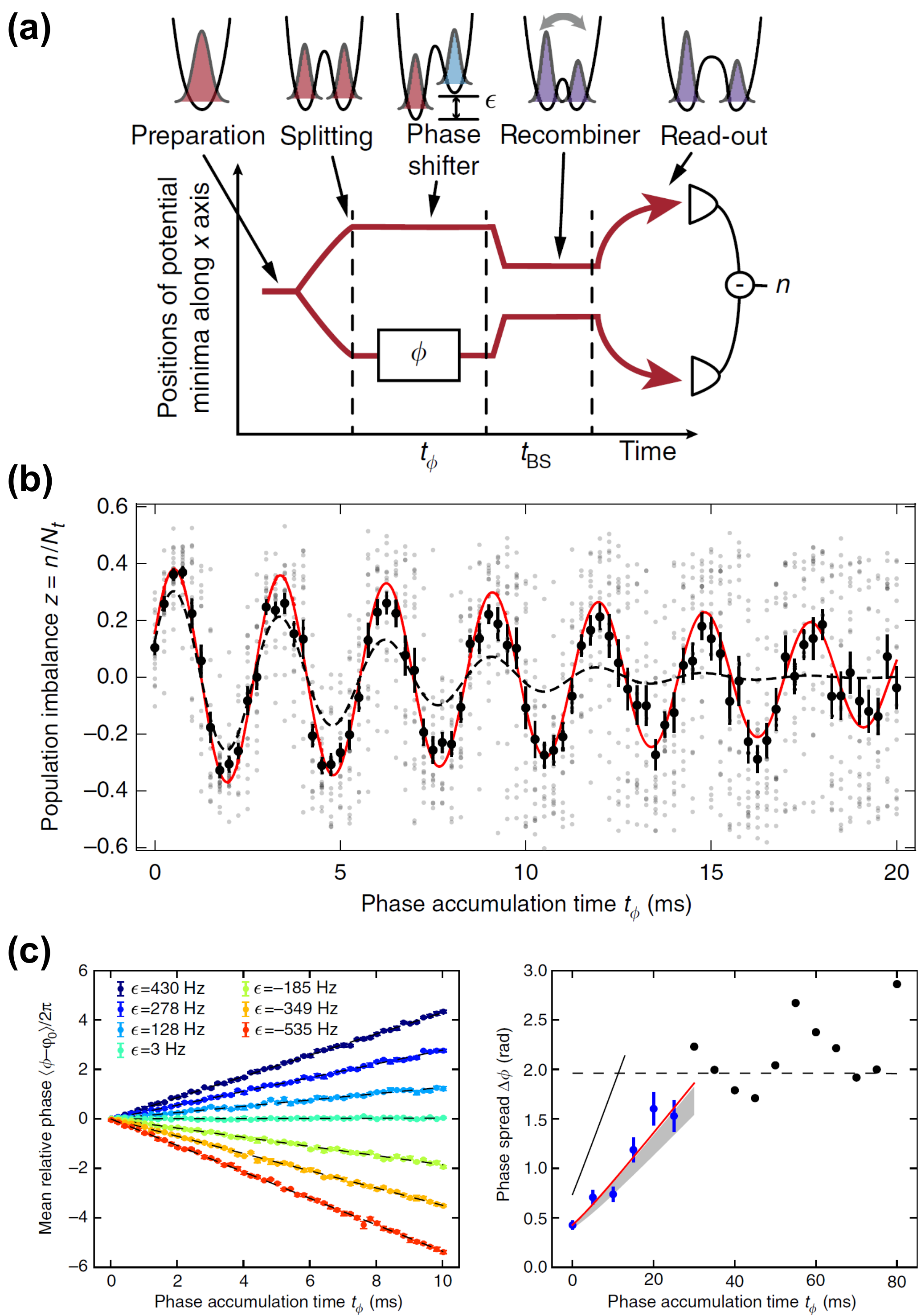}
    \caption{Mach-Zehnder interferometer. a) Schematic illustration of the Mach-Zehnder interferometer on an atomchip. The first beamsplitter is implemented via double-well splitting; tunneling dynamics during the recombineer phase maps the relative phase onto an atom number imbalance. b) Population imbalance at the interferometer output for different phase accumulation times. c) Mean  phase accumulation versus hold time for different energy shifts. d) Phase spread due to atom number fluctuations; the dephasing can be reduced by introducing relative atom number squeezing (for detailed analysis see \cite{Zhang2024}). Taken and adapted from \cite{Berrada2013}.}
    \label{fig:atom_chip_Mach_Zehnder}
\end{figure}

A complementary realization is the {Mach--Zehnder--type interferometer} on an atom chip, which follows a more controlled sequence of beam splitting, phase accumulation, and recombination. After the initial splitting, the two condensates are held in separate wells, where they accumulate a relative phase at a rate determined by the energy offset between the wells. During this interrogation time, interactions can cause self-interaction--induced phase diffusion when there is population imbalance. However, by employing number-squeezed states generated via repulsive interactions (see Section~\ref{section:coherent_splitting}), this phase diffusion can be substantially suppressed, leading to extended coherence times and enhanced phase stability. The mean relative phase increases linearly with the holding time, demonstrating precise control over the system’s evolution. Recent experiments further report spatially resolved phase correlations and coherence measurements in such configurations~\cite{Zhang2024}.

Together, these two interferometric schemes illustrate the versatility of atom-chip platforms for both {quantum sensing} and {quantum simulation}. The double-well interference approach emphasizes interaction-driven many-body dynamics, while the Mach--Zehnder configuration provides a route toward high-precision, controllable matter-wave interferometry.

\section{Interferometers with trapped strongly interacting Feshbach molecules} 
\label{section:Li_molecule_interferometers}

\begin{figure}
    \includegraphics[width = \columnwidth]{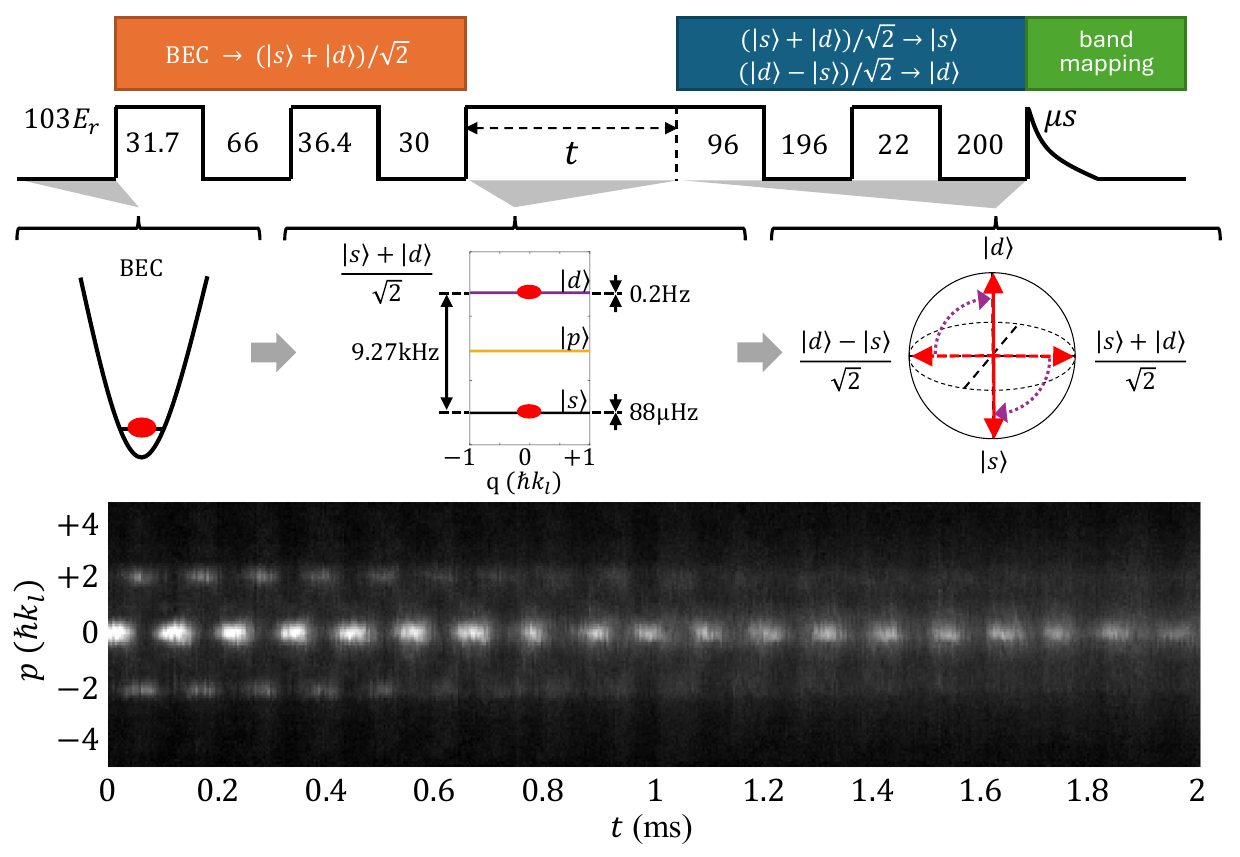}
    \caption{Experimental schemes of the trapped Ramsey interferometer realized with strongly interacting $Li_2$ molecules, and examples of observed interferences. Adapted from \cite{Li2024}.}
    \label{fig:Li_Ramsey_interferometer}
\end{figure}

Strongly interacting Feshbach molecules provide a complementary platform for matter-wave interferometry in which the interaction energy is comparable to, or even exceeds, the recoil and lattice energy scales. 
Short diffraction pulses of an optical lattice---introduced in Sec.~\ref{section:Li_molecule_diffraction}---serve as beam splitters and recombiners, enabling the coherent preparation and manipulation of molecular wave packets in different motional states. 
Depending on the pulse sequence, two distinct trapped interferometer geometries can be realized: 
(i) a Ramsey-type interferometer based on a superposition of lattice-band states (Fig.~\ref{fig:Li_Ramsey_interferometer}), and 
(ii) a Michelson-type interferometer based on momentum-split wave packets propagating in opposite directions within a weak waveguide (Fig.~\ref{fig:Li_Michelson_interferometer}).  
These schemes allow us to examine how strong interactions influence phase evolution, contrast, and coherence in molecular interferometry.

\subsection{Ramsey-type interferometer}

The first scheme uses the optical lattice to couple the molecular condensate into a coherent superposition of the ground band $\ket{s}$ and the second excited band $\ket{d}$, forming an effective two-level motional system~\cite{Hu_2018_Ramsey_motional}. 
A pair of short lattice pulses implements a $\pi/2$ beam splitter, preparing the state $(\ket{s}+\ket{d})/\sqrt{2}$. 
During a variable hold time $t$ in the static lattice, the $\ket{s}$ and $\ket{d}$ components accumulate a relative phase determined by the band-gap energy and by interaction-induced mean-field shifts. 
A second $\pi/2$ pulse maps the accumulated phase onto measurable population differences, which are read out using band mapping~\cite{Greiner_2001_2D_lattice}. 
The populations show damped oscillations as a function of $t$, and fitting these fringes reveals how strong interactions reshape the interferometer.

\begin{figure}
\center
\includegraphics[width=0.6\columnwidth]{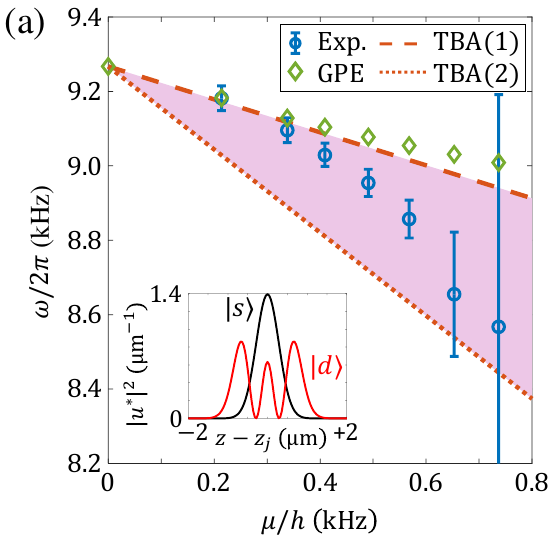}
\includegraphics[width=\columnwidth]{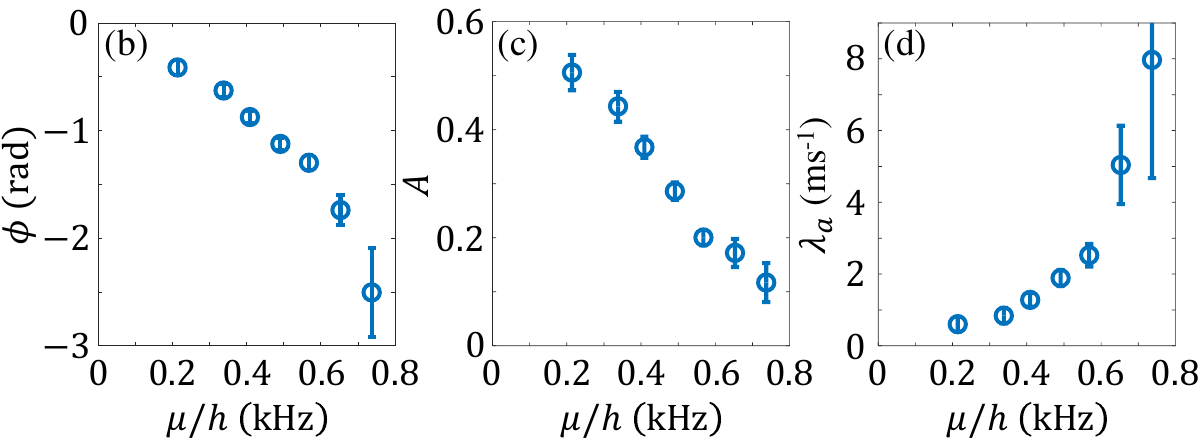}
\caption{Interaction effects in the Ramsey interferometer: (a) Interference frequency,
(b) phase shift, (c) maximal contrast, and (d) contrast decay rate as functions of the chemical potential of the molecular condensate prior to lattice loading, which characterizes the interaction strength. In (a), the interference frequencies fall within the shaded area between two lines obtained by calculations of the energy shift due to interaction under the tight-binding approximation (TBA)}
\label{fig:ramsey_aVar2}
\end{figure}

Figure~\ref{fig:ramsey_aVar2} summarizes the interaction effects:
(i) The initial beam-splitter operation introduces an interaction-dependent phase shift between $\ket{s}$ and $\ket{d}$, which increases with chemical potential (Fig.~\ref{fig:ramsey_aVar2}(b)). 
The pulse sequence is designed to yield zero differential phase in the non-interacting limit, and extrapolation of the data towards vanishing interaction indeed approaches zero.
(ii) Interparticle interaction reduces the interference frequency by modifying the $\ket{s}$--$\ket{d}$ energy gap. 
Because the higher-band state is spatially more extended, its mean-field energy shift is smaller, causing the band gap---and thus the Ramsey frequency---to decrease with increasing interaction. 
A mean-field calculation within a tight-binding approximation reproduces the observed frequency shifts (shaded region in Fig.~\ref{fig:ramsey_aVar2}(a)).
(iii) The initial contrast decreases with interaction strength, and the contrast decay rate increases, reflecting interaction-induced dephasing between the two motional states (Fig.~\ref{fig:ramsey_aVar2}(c) and (d)).

These observations demonstrate that even a simple trapped Ramsey sequence becomes an interaction-sensitive probe of molecular mean-field shifts, coherence, and nonlinear phase evolution.

\subsection{Michelson-type interferometer}

In the Michelson-type interferometer, lattice pulses split the molecular condensate into two counterpropagating momentum components with momenta $\pm 2\hbar k_L$. 
The weak axial confinement of the optical dipole trap acts as a waveguide in which the two wave packets move in opposite directions, accumulate a differential phase, and return to the trap center after approximately half an oscillation period. 
A magnetic-field gradient pulse may be applied to imprint a controlled phase difference between the arms. 
A subsequent lattice pulse recombines the two wave packets, and the resulting momentum distribution gives the interferometric signal.

The strongly interacting character of Li$_2$ molecules manifests itself in several ways:
(i) Imperfect splitting and recombination appear as small occupations in the $\ket{\pm 4k}$ diffraction modes, reflecting sensitivity of the light-pulse operations to interaction energy. 
(ii) As interaction strength increases, collisional loss becomes the dominant decoherence mechanism. 
If recombination is attempted after a \emph{full} oscillation within the waveguide rather than at the first returning to the center, the wave packets cross once more, and the contrast can drop to $\sim20\%$.
(iii) Despite these challenges, clear interference fringes are observed—even for thermal samples—demonstrating the robustness and ``white-light'' character of the Michelson geometry.

\begin{figure}
    \includegraphics[width = \columnwidth]{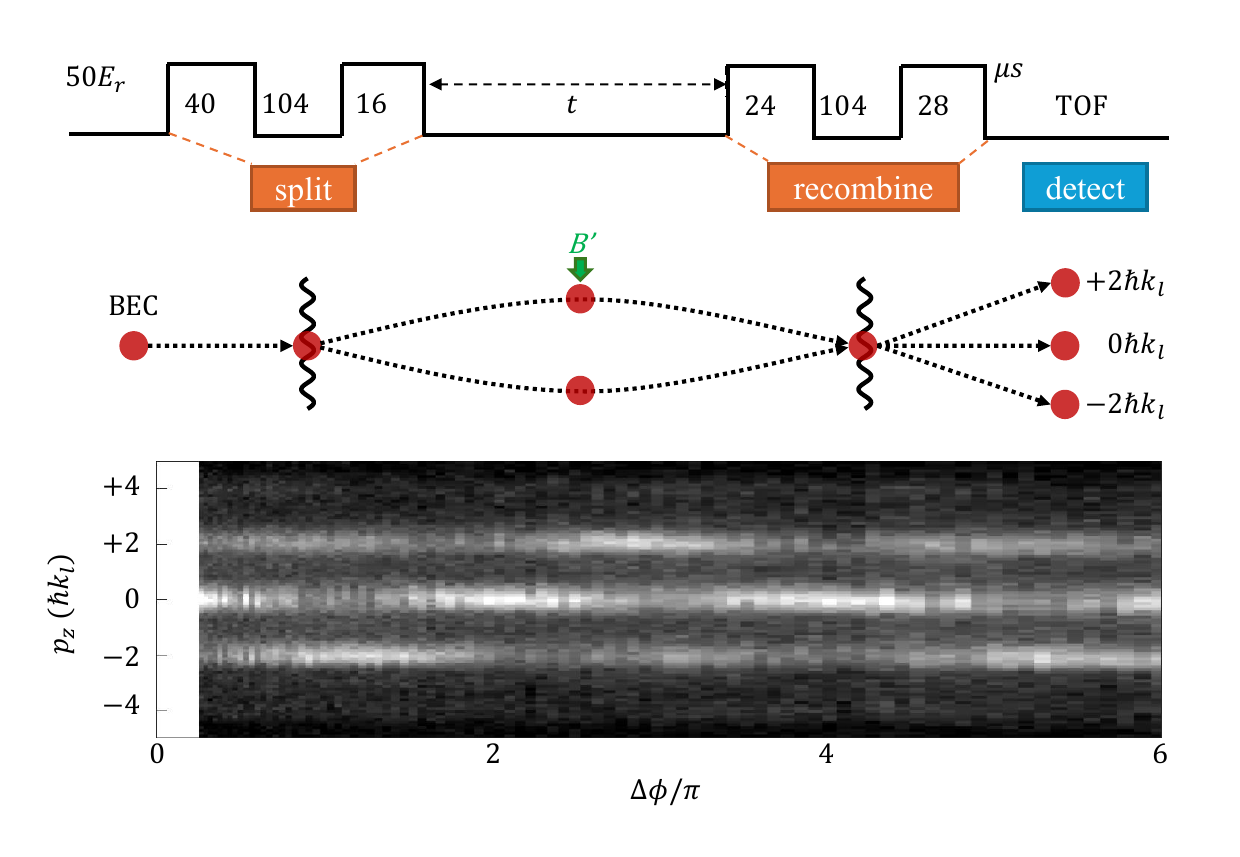}
    \caption{Experimental schemes of the trapped Michelson interferometer realized with strongly interacting $Li_2$ molecules, and examples of observed interferences. Adapted from \cite{Li2024}.}
    \label{fig:Li_Michelson_interferometer}
\end{figure}

Together, the Ramsey and Michelson schemes highlight how strong interactions affect all stages of a molecular interferometer: the beam-splitter operation, the phase-accumulation time, and the recombination contrast.

In the experiments of Refs.~\cite{Liang2022,Li2024}, the condensates typically contain $N\!\sim\!10^4$ Li$_2$ molecules. 
Although the moderate axial confinement keeps peak densities relatively low ($\sim 3 \times 10^{11}\,\mathrm{cm^{-3}}$), the mean-field interaction energy 
$E_{\mathrm{int}} = g n$ with $g = 4\pi\hbar^2 a_s/m$ 
remains comparable to the lattice recoil energy $E_{\mathrm{rec}}$. 
Tuning the scattering length via the Feshbach resonance allows the system to access a regime where interactions dominate even the most elementary optical elements of the interferometer. 
These molecular interferometers therefore provide a unique window into nonlinear matter-wave optics in the strongly interacting regime.

\section{Conclusion and outlook}

Matter-wave optics with interacting quantum gases reveals that even the most elementary optical elements—diffraction gratings, beam splitters, and interferometers—acquire nonlinear characteristics once interparticle interactions are significant. 
Across the examples discussed here, from the diffraction of strongly interacting Feshbach molecules to the coherent splitting and interference of condensates on an atom chip, we observe how interactions transform simple linear processes into rich many-body dynamics. 
Interactions not only modify phase evolution and coherence but also enable new functionalities such as squeezing, entanglement, and engineered nonlinear responses.

Looking ahead, these advances open a pathway towards a comprehensive framework of nonlinear matter-wave optics, paralleling the development of nonlinear photonics. 
Future experiments will exploit tunable interactions, multi-mode geometries, and hybrid coupling to optical or microwave fields to explore novel regimes of quantum transport, correlated light–matter dynamics, and entanglement generation. 
Such studies will bridge quantum metrology, many-body physics, and quantum simulation, extending the legacy of matter-wave interferometry into the strongly correlated domain.

\section*{Acknowledgements}

 We would like to thank the members of the Vienna (former Heidelberg) quantum experiment team and our large number of theory collaborators. This perspective is based on their amazing experiments and insights accumulated over many years.
 This research was supported by the European Research Council: ERC-AdG ``Emergence in Quantum Physics'' (EmQ) under Grant Agreement No. 101097858 and the DFG/FWF CRC 1225 ``ISOQUANT'', with the Vienna participation financed by the Austrian Science Fund (FWF) [grant DOI: 10.55776/I4863] and the Austrian Science Fund (FWF) 'NEqD-si1D' [DOI: 10.55776/P35390]. M. P. has received funding from Austrian Science Fund (FWF) [Grant DOI: 10.55776/ESP396] (QuOntM).

\bibliography{Bibliography}

\end{document}